\documentclass[aps,prd,twocolumn,nobibnotes,nofootinbib,showpacs]{revtex4-1} 

\usepackage{times}
\usepackage{amsmath}
\usepackage{amssymb}
\usepackage{graphicx}
\usepackage{color}
\usepackage[colorlinks]{hyperref}

\definecolor{CiteColor}{rgb}{0,0.5,0}
\hypersetup{citecolor=CiteColor}
\definecolor{RefColor}{rgb}{0.55,0,0}
\hypersetup{linkcolor=RefColor}

\newcommand{\beq}{\begin{equation}}
\newcommand{\eeq}{\end{equation}}
\newcommand{\bsubeq}{\begin{subequations}}
\newcommand{\esubeq}{\end{subequations}}

\renewcommand{\vec}[1]{\ensuremath{\mathbf{#1}}}

\begin{document}

\title{QED Plasma and Magnetars}

\author{Marat Freytsis and Samuel E. Gralla}

\affiliation{Center for the Fundamental Laws of Nature, Harvard University,
Cambridge, MA 02138, USA}

\begin{abstract}
Magnetars are surrounded by diffuse plasma in magnetic field strengths well above the quantum electrodynamic critical value. We derive equations of ``quantum force-free electrodynamics'' for this plasma using effective field theory arguments. We argue that quantum effects do not modify the large scale structure of the magnetosphere, and in particular that the spin-down rate does not deviate significantly from the classical result. We provide definite evolution equations that can be used to explore potentially important small-scale corrections, such as shock formation, which has been proposed as a mechanism for both burst and quiescent emission from magnetars.
\end{abstract}

\maketitle

\textit{\textbf{Introduction ---}} 
From the earliest days of the quantum theory of light, before even the development of quantum electrodynamics (QED) proper, it was recognized that quantum effects should become important for electromagnetic field strengths of order $m^2/\hbar e$, where $m$ and $e$ are the mass and charge of the electron~\cite{Sauter:1931zz,Heisenberg:1935qt}.  New effective photon-photon interactions emerge, mediated by electron loops, leading to phenomena such as vacuum birefringence and light-by-light scattering.  Most dramatically, critical-strength electric fields create electron-positron pairs out of the vacuum, a non-perturbative effect~\cite{Schwinger:1951nm}.  The most promising route to reaching these field strengths in the laboratory is the use of high-intensity lasers \cite{DiPiazza:2011tq}. While some of the effects may be observable in upcoming facilities, the actual field strengths will still be subcritical.

Fortunately, nature provides us with another avenue to investigate strong-field QED: a class of astrophysical objects known as \textit{magnetars}. Magnetars are pulsars (rotating, magnetized neutron stars) with exceptionally strong surface magnetic field strengths.  In fact, magnetars can have field strengths of up to $10^{15} \text{ G}$ and maybe higher, which exceed the critical field strength, 
\begin{equation}\label{crit}
B_Q = \frac{m^2}{\hbar e} \approx 4.4 \times 10^{13} \text{ G},
\end{equation}
by two orders of magnitude!  Much work has been devoted to understanding the physical processes that take place in such magnetic field strengths; see \cite{Lai:2014nma,Harding:2006qn,Uzdensky:2014rza} for reviews.

Most of this work is done assuming a vacuum environment, whereas in fact magnetars (and pulsars in general) are surrounded by a diffuse plasma.  The existence and properties of this plasma can be understood from the smallness of the dimensionless parameter
\begin{equation}\label{hummingbird}
  \chi = \frac{m}{eBR} \approx 10^{-15}.
\end{equation}
Here $B$ is the magnetic field strength, $R$ is the stellar radius, and we have assumed canonical pulsar values $B \approx 10^{12} \text{ G}$ and $R \approx 10 \text{ km}$.

This number accounts for the plasma as follows \cite{Goldreich:1969sb,Ruderman:1975ju}.  A conductor moving with velocity $v$ in a magnetic field $B$ generates an electric field of order $B v$ by ``unipolar induction''.  For a rotating magnetized sphere in vacuum this electric field has a component along $B$, and hence can accelerate particles.  The energy to which the particles can be accelerated over a typical system size is thus $e B v R$.    Computing $v/\chi$ shows that this energy exceeds the rest mass of the electron by many orders of magnitude.  (For pulsars a typical surface velocity is $v\sim10^{-4}$.)  Thus any stray charges are rapidly accelerated to above the pair-production threshold, and the ensuing pair-creation cascade will populate the magnetosphere with plasma.

As charges are generated they will arrange themselves to cancel the electric field, driving the Lorentz invariant $\vec{E} \cdot \vec{B}$ to zero.  Production ceases as this invariant becomes too small to produce the required acceleration.  For charge corotating with the star the density required to cancel $\vec{E} \cdot \vec{B}$ is the so-called Goldreich--Julian charge density $vB/R$.  This sets a typical scale for the plasma mass density, $mvB/eR$.   The ratio of the particle mass/energy density to the electromagnetic field energy density is then $v \chi$, which is exceedingly tiny, making the plasma dynamics completely dominated by the field.

Assuming classical electrodynamics, such plasmas are described by a non-linear theory of the electromagnetic field known as force-free electrodynamics (FFE) \cite{Goldreich:1969sb,uchida1997general,Gruzinov:1999aza,Gralla:2014yja}.  The theory follows from the assumption that the electromagnetic stress-energy is conserved on its own, leading to the condition that the Lorentz force density everywhere vanishes. Naively, one might expect any classical description to break down at or near the critical field strength $B_Q$.  The measurement of the surface magnetic field strength of a pulsar/magnetar relies on the dipole radiation spin-down formula, which has only been derived in classical electrodynamics (vacuum~ \cite{1955AnAp...18....1D,Rezzolla:2000dk,Rezzolla:2004hy} or force-free~\cite{Contopoulos:1999ga,Spitkovsky:2006np,Ruiz:2014zta}) or in vacuum QED~\cite{Heyl:1997hs}.  Thus the very evidence for super-critical magnetic fields in nature is sensitive to the question of magnetically dominated QED plasma.  

In this letter we will derive equations of ``quantum force-free electrodynamics'' to describe this plasma.  The strategy is to integrate out electron loop fluctuations from the QED action, following the basic approach established by Euler and Heisenberg (EH) in 1935~\cite{Heisenberg:1935qt}.  However, the EH calculation is done assuming no fermion in- or out-states, allowing the electron to be integrated out entirely in the effective action, whereas we wish to consider plasma.  We therefore proceed in two steps.  First, we consider a collisionless multiparticle system and use effective field theory (EFT) arguments to establish the size and form of the modifications due to QED.  We then show that the modifications that survive in the magnetically dominated limit follow from the EH Lagrangian, without requiring a new QED calculation.  We thereby write down definite equations for magnetically dominated QED plasma.

Like their classical counterpart, the quantum force-free equations imply $\vec{E}\cdot{\vec{B}}=0$.  This makes the EH Lagrangian real and pole-free (no Schwinger pair production), which implies that the effective theory will adequately describe physics at and above the critical field strength.  The classical force-free equations are supplemented by small corrections, and there is no drama at the critical scale for magnetars.   In particular, we expect the classical spindown formula to be corrected only by a factor of at most order $10^{-3}$, supporting the self-consistency of the argument for strong magnetic fields in magnetars.  

Small corrections can still have important effects if they give rise to qualitatively new features.  One new feature here is the non-linearity of the vacuum sector of the theory, which gives rise to shock formation (e.g., \cite{kovachev-georgieva-kovachev2012}).  Heyl and Hernquist have argued that such shocks also form in the plasma context \cite{Heyl:1998uw,Heyl:1998db} and that the energy in the shocks and would eventually be dissipated into electron-positron pairs, providing a mechanism for both burst \cite{Heyl:2003wn} and quiescent  \cite{Heyl:2005an,Heyl:2006uy} magnetar emission.  Their results are based primarily on an analysis of the characteristics of the one-dimensional, linearized problem.  Our new, non-linear evolution equations for the full three-dimensional description of the quantum plasma provde a foundation for exploration of the detailed operation of this and other QED effects in realistic magnetosphere models.

It is noteworthy that a relativistic QED plasma can be given such a simple description in the strong-field limit.  We show how this simplification occurs for any covariant Lagrangian assumed to have a standard matter coupling and to have conserved stress-energy without matter contributions.  We discuss how every Lorentz-invariant action gives rise to a ``force-free'' dynamics in addition to its ordinary least-action dynamics.

We use Heavisde--Lorentz units with the speed of light $c=1$, but leave $\hbar \neq 1$.  Our metric signature is $(-,+,+,+)$.  We restrict to flat spacetime, but all results generalize straightforwardly to an arbitrary curved spacetime.

\textit{\textbf{Particle EFT ---}} 
We begin by considering a system of $N$ point particles coupled to a gauge field, in the spirit of the analyses carried out in Refs.~\cite{1975ApJ...196L..59E,Goldberger:2004jt}.  We will assume that these particles do not collide, which is justified by the low density of the magnetar plasma.  This means that the EFT action will be a functional the gauge field $A_\mu$ together with $N$ worldlines $z^\mu_i$, with all terms local, and gauge, Lorentz, and worldline-reparameterization invariant. If supplemented with worldline spinors coupling to the gauge field, this formulation is fully equivalent to ordinary QED~\cite{Brink:1976uf,Strassler:1992zr}, except that the restriction to fixed $N$ limits its validity to processes where electron-position pairs cannot be created.  In the derivative expansion, all effects of the spinor are parametrized by their contributions to multipole couplings of the particle.

As in the derivation of the standard EH Lagrangian, the power-counting parameter is the average momentum transfer over the particle mass, so that derivatives are suppressed by the momentum of the Fourier transform of the field strength over the electron mass. In magnetar applications this is the electron Compton wavelength over the neutron star radius, justifying keeping only to leading non-trivial order in the derivative expansion.  In practice this means that only terms made of $A_{\mu}$, $F_{\mu \nu}$, and the 4-velocity $u_i^\mu$ may appear, and we integrate along the worldlines with respect to proper time $\tau_i$.  The complete collection of terms is\footnote{Dipole couplings also appear at this order in the derivative expansion, and technically should be included in Eq.~\eqref{Seff}.  However, after matching to QED and taking the magnetically dominated limit these terms are the same order as those we drop.  For simplicitly we do not write them down at this stage.}
\begin{multline}\label{Seff}
  S_{\rm eff} = \int \mathcal{L}_{\rm EM}[F_{\mu \nu}] d^4 x \\
    - \sum_{i=1}^N \int_{z_i} \left(m_i[F_{\mu \nu},u_i^\mu] - q_i \ u_i^\mu A_\mu\right) d\tau_i.
\end{multline}
Here $q_i$ is a constant interpreted as the particle charge.  If $m_i$ were likewise constant, it would be the standard notion of particle mass.  However, here we say only that $m_i$ is a function with dimensions of mass that is constructed covariantly and gauge-invariantly from $F_{\mu \nu}$ and $u^\mu_i$.  Finally $L_{\rm EM}$ is an arbitrary function of $F_{\mu \nu}$ only, and hence can depend only on the invariants
\begin{equation}
  I = F_{\mu \nu} F^{\mu \nu}, \quad
  K = \tilde{F}_{\mu \nu} F^{\mu \nu}, \quad
  \tilde{F}_{\mu \nu}=\tfrac{1}{2} \epsilon_{\mu \nu \rho \sigma} F^{\rho \sigma}.
\end{equation}
In $(3+1)$ langauge these invariants are $I=2(B^2-E^2)$ and $K=2\vec{E} \cdot \vec{B}$.  The ``mass'' $m_i$ can depend on these invariants as well as the third invariant $F_{\mu \alpha} F^\mu{}_{\beta}u^\alpha u^\beta$.

Varying with respect to the gauge field $A^\mu$ gives
\begin{align}\label{eom-A}
  -4\nabla_\nu \left\{\frac{\partial L_{\rm EM}}{\partial I}  F^{\mu \nu}
     + \frac{\partial L_{\rm EM}}{\partial K} \tilde{F}^{\mu \nu} \right\} = J^\mu
\end{align}
with
\begin{align}\label{eom-J}
  J^\mu = \sum_{i=1}^N \int_{z_i} \left(q_i u^\mu
            + 2\nabla_\nu \frac{\partial m_i}{\partial F_{\mu \nu}}\right) \delta(x-z_i(\tau)) d\tau_i.
\end{align}
Varying with respect to $z^\mu$ gives for each particle $i$,\footnote{Since $\tau_i$ depends on the worldline, one should instead vary with respect to $z_i^\mu(\lambda)$ for some fixed parameter $\lambda$.}
\begin{multline}\label{eom-Z}
  \frac{d}{d\tau_i}
    \left\{ m_i u_i^\mu + (g^{\mu \nu}+u_i^\mu u_i^\nu)\frac{\partial m_i}{\partial u_i^\nu} \right\} \\
    = q_i  F_\nu{}^\mu u_i^\nu - \frac{\partial m_i}{\partial F_{\alpha \beta}} \nabla_\mu F_{\alpha \beta}
\end{multline}
As with ordinary electromagnetism coupled to point particles, Eqs.~\eqref{eom-A}-\eqref{eom-Z} mix $\delta$-functions with nonlinearities and hence must be regarded as formal.  The origin is that the EFT should not describe physics on the scale of the particle size.  Restricting to the appropriate larger scales, each particle makes a small perturbation to the long-wavelength field produced by the other particles, and the delta functions appear as point sources for the linearized version of Eq.~\eqref{eom-A}.  Equivalently, one can think of the force law \eqref{eom-Z} as holding for the field produced by all other charges except the one under consideration.  The neglected self-force effects are small in the magnetar application, but may be important for laser plasmas \cite{DiPiazza:2011tq}.

\textit{\textbf{QED Matching and the Magnetic Limit} ---} The unknowns $q_i$, $m_i$, and $\mathcal{L}_{\rm EM}[I,K]$ are to be matched with a UV theory valid on small scales, in this case QED.  The charges $q_i$ match trivially with $\pm e$ (with $+$ for positrons and $-$ for electrons) based on the form of the interaction term in the QED Lagrangian.  To determine $L_{\rm EM}$ we note that this term does not depend on $u_i^\mu$ and hence can be considered in the special case $N=0$ where there are no particles in the EFT.  On scales smaller than the typical variation $R$ of the field the field strength $F_{\mu \nu}$ is approximately constant.  Thus we want to match to QED in the presence of a uniform external field.  Since there are no fermion degress of freedom in the EFT we must integrate these out of QED.  This is the classic Euler--Heisenberg calculation, and hence
\begin{equation}
\mathcal{L}_{\rm EM} = \mathcal{L}_{\rm EH},
\end{equation}
where the expression for $\mathcal{L}_{\rm EH}$ may be found in standard texts (e.g., Eq.~(1.2) of Ref.~\cite{Dunne:2004nc} or Ch.\ 33 of Ref.~\cite{Schwartz:2013pla}). 

The functional form of the ``mass'' $m_i[F_{\mu \nu},u^\mu]$ will be presented in a future paper.\footnote{This term gives rise to strong-field corrections to the Lorentz force law, of the variety considered in Ref.~\cite{Labun:2013lza,1996PhRvE..54..884W}.}  For the present paper we will need only the fact that $m_i$ is a real and slowly-growing function of field strength in the magnetic case $I>0, \ K=0$ of relevance here, which follows from results already known in the literature.  In particular, the electron ground state energy shift in a strong magnetic field~\cite{Jancovici:1970ep,Constantinescu:1972qe} is order $\alpha m$ in all interesting magnetic field strengths.  There are also corrections to $m_i$ proportional to the velocity $u^\mu$, but these follow from matching to the same calculation in the UV theory, namely, the renormalization of the exact electron propagator in a background field, and so must have the same analytic structure.

The (related) facts that the magnetar field is magnetic ($I>0$) and the effective mass is order $m$ turn out to allow us to drop all terms involving $m_i$ for that application. For example, in Eq.~\eqref{eom-J} the spacetime derivative counts as a power of inverse length $1/R$, while the field strength derivative counts as a power of $1/B$.  Thus dividing the second term by the first, the ratio $m/eBR$ is the parameter $\chi$ defined earlier in Eq.~\eqref{hummingbird}, and the current is given to an excellent approximation as 
\begin{align}\label{eom-J2}
  J^\mu = \sum_{i=1}^N \int_{z_i} q_i u^\mu \delta(x-z_i(\tau)) d\tau_i.
\end{align}
The mass terms in the force law \eqref{eom-Z} are also formally supressed in this way, but dropping them requires more care, and relies crucially on the field being magnetic, rather than electric or null.  The subtlety is that the LHS of Eq.~\eqref{eom-Z} contains terms of higher differential order, which therefore can have their scale set by the dominant (Lorentz) force together with initial conditions.  In the magnetic case this scale is the cyclotron motion about the field line, and dropping $m$ from the Lorentz force law averages over the gyrations \cite{kruskal1958,northrop1961}. (In strong fields the cyclotron motion is quantized, so energy eigenstates don't show any actual gyration.)  Setting $m_i=0$ in Eq.~\eqref{eom-Z}, we arrive at
\begin{equation}\label{eom-Z2}
  F_{\mu \nu} u_i^\nu = 0.
\end{equation}
The content of this statement is that the particles are stuck to magnetic field lines, moving freely along them in the guiding center approximation.  For further discussion of the relativistic meaning of this statement, see  Sec.~3.2.3 of \cite{Gralla:2014yja}.

The relationship between the small parameter $\chi$ and the EFT \eqref{Seff} is clarified by rewriting the former as
\begin{equation}\label{chalternate}
  \chi = \frac{m}{eBR} = \frac{B_Q}{B} \frac{\hbar}{mR}.
\end{equation}
The second factor $(\hbar/m)/R$ is the power-counting parameter of the derivative expansion. Thus $\chi$ being small in large magnetic fields automatically implies the worldline EFT is valid.  On the other hand, the EFT can still be valid when $\chi$ is large, such as if the fields are nearly null ($B=\sqrt{I/2}\ll B_Q$) but still strong, $F_{\mu\nu} \sim B_Q$.  The derivation above emphasizes the separate physical origin of the quantum corrections from the mass-independence of the particle equations of motion, highlights the role of the magnetic assumption in the latter, and avoids subtleties with the orders of limits involved in keeping $m$ finite to derive the EH terms, while setting $m=0$ in the particle worldlines.

\textit{\textbf{Quantum FFE ---}} Dotting both sides of Eq.~\eqref{eom-J2} with $F_{\mu \nu}$ and using Eq.~\eqref{eom-Z2} gives the \textit{force-free} condition,
\begin{equation}\label{force-free}
F_{\mu \nu}J^\nu=0.
\end{equation}
This is the statement that the Lorentz force density vanishes at every point in the plasma.  Using \eqref{eom-A} to eliminate $J^\mu$ in favor of $F_{\mu \nu}$, we obtain a closed system of equations for the electromagnetic field, without reference to the charges.  (The charges move on field lines to provide the current, but their mass does not affect the field dynamics.)  

The existence \eqref{force-free} of a zero-eigenvector $J^\mu$ for $F_{\mu \nu}$ implies that the second invariant vanishes, $K=0$.  Such fields are called degenerate and have a beautiful mathematical structure \cite{carter1979,uchida1997general,Gralla:2014yja}.  They define a foliation of spacetime by two-surfaces, which in the magnetic case $I>0$ are timelike and represent worldsheets of magnetic field lines, or \textit{field sheets}.  They can always be expressed as the wedge of two one-forms $F_{\mu \nu}=a_{[\mu}b_{\nu]}$, which themselves are exact, $a_\mu=\nabla_\mu \phi_1$ and $b_\mu=\nabla_\mu \phi_2$ for two scalar ``Euler potentials'' $\phi_1$ and $\phi_2$.

In light of the degeneracy of $F_{\mu \nu}$ we may now simplify the force-free equations.  Substituting Eq.~\eqref{eom-A} into Eq.~\eqref{force-free} and using $F_{\mu \nu}=a_{[\mu}b_{\nu]}$ it follows that the term involving $\tilde{F}$ always vanishes, and we obtain
\begin{equation}\label{ffe-gen1}
F_{\mu \nu} \nabla_\rho \left\{ f(I) F^{\rho \nu} \right\} = 0, \quad  f\equiv \left. -4\frac{\partial L_{\rm EM}}{\partial I} \right|_{K=0}. 
\end{equation}
Using the explicit form of the EH Lagrangian, we have
\begin{equation}\label{f}
f = 1 + g + O(\alpha^2),
\end{equation}
with
\begin{equation}\label{g}
  g(I) = \frac{4\alpha}{\pi} \int_0^\infty \frac{ds}{s^2}\left(1 + \frac{s}{2b}\right)
           \left( \coth s - \frac{1}{s} - \frac{s}{3} \right)e^{-s/b}.
\end{equation}
Here we introduce the dimensionless magnetic field strength $b=B/B_Q=\sqrt{I/2}/B_Q$. The integral in \eqref{g} is finite and can be computed numerically at a given value of $b$ or expanded at strong or weak fields \cite{Dunne:2004nc}.  It is of order $10^{-3}$ at magnetar field strengths.  This expression includes all terms involving one fermion loop.  It is corrected at $O(\alpha^2)$ by higher loops, but is non-perturbative in the field strength $b$, holding at arbitrarily\footnote{Since $\mathcal{L}_{\rm EM}$ grows only like $\alpha \log b$ at large $b$, there is no reason to suspect a breakdown of perturbation theory until the fantastical value of $b\sim e^{137}$, corresponding to field strengths of $\sim 10^{60}G$.} strong fields. Eqs.~\eqref{ffe-gen1}, \eqref{f} and \eqref{g} are the theory of \textit{quantum force-free electrodynamics} in the one-loop approximation.

\textit{\textbf{Evolution form ---}} While the elegant form \eqref{ffe-gen1} is convenient for general manipulations, for numerical solution one must express the theory in the form of evolution and constraint equations.  We take the electric and magnetic fields $\vec{E}$ and $\vec{B}$ as our fundamental variables. We begin by defining a \textit{fictitious} current $\bar{J}^\mu$ by
\begin{equation}
\bar{J}^\mu \equiv \nabla_\nu [ f(I) F^{\mu \nu} ],
\end{equation}
in terms of which the equations of motion \eqref{ffe-gen1} still take the form $F_{\mu \nu}\bar{J}^\mu=0$. In this form it is clear that the inhomogeneous Maxwell equations are modified by $\vec{E} \rightarrow f \vec{E}$, $\vec{B} \rightarrow f \vec{B}$, $\rho \rightarrow \bar{\rho}$ and $\vec{J} \rightarrow \vec{\bar{J}}$,
\begin{align}
\vec{\nabla} \cdot (f \vec{E}) & =\bar{\rho} \label{biscuit1} \\
\vec{\nabla} \times (f \vec{B}) & = \vec{\bar{J}} + \partial_t ( f \vec{E} ). \label{biscuit2}
\end{align}
The homogeneous Maxwell equations $\nabla_{[\mu}F_{\rho \sigma]}=0$ are unmodified,
\begin{align}
  \vec{\nabla} \cdot \vec{B} & = 0 \label{biscuit3} \\
  \vec{\nabla} \times \vec{E} & = -\partial_t \vec{B}. \label{biscuit4}
\end{align}
The force-free condition $F_{\mu \nu}\bar{J}^\mu=0$ is equivalent to
\begin{align}
  \vec{E} \cdot \vec{\bar{J}}=0, \quad \bar{\rho} \vec{E} + \vec{\bar{J}}\times \vec{B}=0. \label{biscuit5}
\end{align}
The fact that $K=0$ for any solution is equivalent to $\vec{E} \cdot \vec{B}=0$, whose time derivative states that
\begin{align}
  \vec{E} \cdot \vec{B} = 0 \quad \rightarrow \quad
  \partial_t \vec{E} \cdot \vec{B} = \vec{E} \cdot (\vec{\nabla} \times \vec{E}), \label{biscuit6}
\end{align}
where \eqref{biscuit4} has been used.

\begin{widetext}
Eq.~\eqref{biscuit4} is the evolution equation for $\vec{B}$.   We determine the evolution equation for $\vec{E}$ by projecting $\partial_t \vec{E}$ onto the basis $\{\vec{E},\vec{B},\vec{E}\times\vec{B}\}$, using Eqs.~\eqref{biscuit1}-\eqref{biscuit6}, and eliminating $\bar{\rho}$ and $\vec{\bar{J}}$.  The result is
\begin{equation} \label{evo1}
    \partial_t \vec{E} = \frac{\vec{E}}{E^2} \left\{ \frac{\vec{E} \cdot \vec{\nabla}\times(f \vec{B})
                          + 4 f' E^2 \vec{B} \cdot \vec{\nabla}\times \vec{E}}{f-4f' E^2}\right\}
      + \frac{\vec{B}}{B^2} \left\{ \vec{E} \cdot \vec{\nabla} \times\vec{E} \right\} 
 + \frac{\vec{E}\times\vec{B}}{E^2 B^2} \frac{1}{f} \left\{\vec{E} \times \vec{B} \cdot \vec{\nabla} \times (f \vec{B}) -E^2 \vec{\nabla} \cdot(f \vec{E}) \right\},
\end{equation}
where a prime represents an $I$-derivative.  If we linearize with respect to $g$ then we find
\begin{multline} \label{evo2}
\partial_t \vec{E} = \vec{\nabla} \times \vec{B} + \frac{\vec{B}}{B^2}(\vec{E} \cdot \vec{\nabla} \times \vec{E} - \vec{B} \cdot \vec{\nabla} \times \vec{B})
 - \vec{\nabla} \cdot \vec{E}\, \frac{\vec{E}\times\vec{B}}{B^2} \\ -  \frac{d g}{d I} \frac{\vec{E} \times \vec{B}}{B^2}\vec{E} \cdot \vec{\nabla} I
+ \frac{d g}{d I} \vec{E} \left[ \frac{\vec{E}\times\vec{B} \cdot \vec{\nabla} I}{E^2} + 4 (\vec{E} \cdot \vec{\nabla}\times\vec{B} + \vec{B} \cdot \vec{\nabla} \times\vec{E}) \right],
\end{multline}
where we remind the reader that $I = 2(B^2-E^2)$.  The first line gives the evolution equation for standard Maxwellian FFE with the last two terms comprising (minus) the Maxwellian current $\vec{J}$.  We may view the remainder of Eq.~\eqref{evo2} providing effective corrections to the current.  The first term of the second line of Eq.~\eqref{evo2} corrects the ``advection'' term $\vec{\nabla} \cdot \vec{E}\, (\vec{E}\times\vec{B}/B^2) = \rho \vec{v}$ (with $\rho$ the charge density and $v$ the drift velocity) for the effective charge density \eqref{biscuit1}.  The second term introduces an effective current directed along the electric field, a qualitatively new feature.  (The real current is orthogonal to the electric field, as required by the force-free equation \eqref{force-free}.)

The evolution system consists of evolution equations $\partial_t \vec{B}=-\vec{\nabla} \times \vec{E}$ and either Eq.~\eqref{evo1} or \eqref{evo2} for $\partial_t \vec{E}$, together with constraints $\vec{\nabla} \cdot \vec{B}=0$ and $\vec{E}\cdot \vec{B}=0$.  By construction, the evolution equations preserve the constraints.  We have not determined whether this formulation is mathematically well-posed. If $g=0$ the sytem is equivalent to ordinary FFE, which can be made hyperbolic provided $I>0$ \cite{Komissarov:2002my,Pfeiffer:2013wza}.
\end{widetext}

\textit{\textbf{Previous Work ---}} 
Heyl and Hernquist postulated the Lagrangian $L_{\rm EH} + \theta K$, where $\theta$ acts as a Lagrange multiplier.  The equations of motion are then $K=0$ as well as Eq.~\eqref{eom-A} together with the expression $J^\mu=\partial_\nu \theta \tilde{F}^{\mu \nu}$ for the current.\footnote{Thompson and Blaes \cite{Thompson:1998ss} have also argued directly for this form of the current based on bozonization of the fermion field on each field line.}  They perturbed from a uniform field and eliminated the Lagrange multiplier in this $(1+1)$D linearized theory.  The Lagrange multiplier may be eliminated in general by recalling that $K=0$ implies  $F_{\mu \nu}=a_{[\mu}b_{\nu]}$ for some (non-unique) pair of one-forms.  Dotting with the above expression for the current gives the force-free condition $F_{\mu \nu}J^\nu=0$.  Conversely, one may show that the force-free condition implies that form for some scalar $\theta$.  Thus the Lagrangian gives rise to the full $(3+1)$, non-linear theory we consider. 

\textit{\textbf{Generalized force-free fields ---}} 
We conclude with a discussion of force-free theories generally. Consider a Lagrangian $L$ assumed only to be a covariant, gauge-invariant functional of the metric and gauge field,
\begin{equation}\label{action-gen}
S = \int  L[g^{\mu \nu}, A_{\mu}] \sqrt{-g} \ \! d^4 x.
\end{equation}
Instead of imposing the equations of motion we define their failure to be satisfied to be the stress-energy and charge-current,
\begin{equation}
T_{\mu \nu} = -\frac{2}{\sqrt{-g}}\frac{\delta S}{\delta g^{\mu \nu}}, \quad J^\mu = -\frac{1}{\sqrt{-g}}\frac{\delta S}{\delta A_\mu}.
\end{equation}
The first is the standard definition of stress-energy, while the second is equivalent to adding a coupling term $A_\mu J^\mu$ to the action \eqref{action-gen}.  The covariance and gauge-invariance of $L_{\rm EM}$ imply that these quantities satisfy (e.g., \cite{Gralla:2010cd})
\begin{equation}\label{ids}
\nabla_\mu J^\mu = 0, \quad \nabla^\nu T_{\mu \nu}=-F_{\mu \nu} J^\nu.
\end{equation}
No field equations are imposed here; these are simply identities that hold for arbitrary vector fields $A^\mu$.  We give the theory content by demanding conservation of stress-energy, $\nabla^\nu T_{\mu \nu}=0$, which immediately implies the force-free condition $F_{\mu \nu}J^\nu=0$.  Allowing non-zero current $J^\mu$ corresponds to allowing unspecified coupling to matter, but demanding conservation of $T_{\mu \nu}$ consists of neglecting the stress-energy of the matter.  This is appropriate when there is much more energy in the field than in the matter.  Thus we provide a simple derivation that the strong-field limit is force-free for an arbitrary classical electrodynamics (such as Born--Infeld).  It would be an efficient derivation of the force-free equations for QED plasma if the coupling $A_\mu J^\mu$ and assumption $\nabla^\nu T_{\mu \nu}=0$ could be justified directly from QED in the appropriate limit.  

This type of calculation need not be restricted to electrodynamic theories, and indeed an identity of the form $\nabla^\nu T_{\mu \nu} = Y[\psi,\delta S /\delta \psi]$ exists for every covariant Lagrangian of some collection of fields $\psi$ \cite{Seifert:2006kv,Gralla:2010cd,Gralla:2013rwa}.  By demanding $\nabla^\nu T_{\mu \nu}=0$ one can define ``force free'' conditions $Y=0$ for any such Lagrangian.  Thus, besides the usual field equations $\delta S /\delta \psi=0$, each covariant Lagrangian gives rise naturally to a second set of equations $Y[\psi,\delta S /\delta \psi]=0$.

\begin{acknowledgements}
We gratefully acknowledge helpful discussions with Clay C\'{o}rdova, Ted Jacobson, Lars Hernquist, Lance Labun, Matt Reece, Ira Rothstein, Matt Schwartz, and Giorgio Torrerei. The work of MF is supported by the Department of Energy under grant DE-SC003916 and the National Science Foundation (NSF) under grant No. PHY-1258729.  S.G. was supported by the NSF under grant No. PHY-1205550. 
\end{acknowledgements}

\bibliography{FFQED}

\end{document}